\documentclass[final] {aipproc}
\usepackage{sidecap,graphicx,boxedminipage,epsfig,wrapfig,floatflt,shadow}

\layoutstyle{8x11double}

\begin{document}

\title{Assessing and improving student understanding of quantum mechanics}


\author{Chandralekha Singh}{
  address={Department of Physics and Astronomy, University of Pittsburgh, Pittsburgh, PA, 15260}
}

\begin{abstract}
We developed a survey to probe student understanding of quantum mechanics concepts 
at the beginning of graduate instruction. The survey was administered to 202 graduate students in physics enrolled in 
first-year quantum mechanics courses from seven different universities at the beginning of the first semester.
We also conducted one-on-one interviews with fifteen graduate students or advanced undergraduate students who had just finished
a course in which all the content on the survey was covered.
We find that students share universal difficulties about fundamental quantum mechanics concepts. The difficulties are often
due to over-generalization of concepts learned in one context 
to other
contexts 
where they are
not directly applicable and difficulty in making sense of the abstract quantitative formalism of quantum mechanics.
Instructional strategies that focus on improving student understanding of these concepts should take into account these difficulties.
The results from this study can 
sensitize instructors of first-year graduate quantum physics to the conceptual difficulties students are likely to face.
\end{abstract}

\maketitle

\section{Introduction}

Quantum physics is a technically difficult and abstract subject. 
The subject matter makes instruction quite challenging, and advanced students constantly struggle to master basic concepts.
Several investigators have explored difficulties in learning quantum mechanics~\cite{zollman,redish,my}.
Here we describe the results from part of a survey of quantum mechanics concepts given to 202 graduate
students from seven different universities.
The 50 minute written survey administered at the beginning of a first-year, first-semester/quarter 
graduate quantum mechanics course covers a wide variety of fundamental quantum concepts. 
To investigate the difficulties with these concepts in depth,
fifteen beginning graduate students or physics seniors from the University of Pittsburgh 
enrolled in an undergraduate quantum mechanics course were interviewed
using a think-aloud protocol~\cite{chi}. 
Students were asked to verbalize their thought processes while they were working on the survey problems.  
Students were not interrupted unless they remained quiet for a while. In the end, we asked them for clarifications of the
issues they had not made clear earlier. 
Although students from some universities performed better on average than others, we find that students have common 
difficulties with many basic concepts in quantum mechanics, regardless of where they are enrolled.

During the design of the survey, we consulted three faculty members at Pitt
who had taught quantum mechanics. Previously, we discussed with them the fundamental concepts that they expected the
students in a first-semester graduate quantum mechanics to know. We iterated the survey several times 
before coming up with the final version. 

\vspace*{-.1in}
\section{Discussion}
\vspace*{-.05in}

Here, we discuss the findings from the part of the survey that covers topics such as
allowed wave functions, the time-dependence of wave functions, 
the probability of measuring energy and the expectation value of energy in a given state. 

\underline{Question (2):} The wave function of an electron in a one-dimensional infinite square well of width $a$ 
at time $t=0$ is given by $\Psi(x,0)=\sqrt{2/7} \phi_1(x)+\sqrt{5/7} \phi_2(x)$ where $\phi_1(x)$ and $\phi_2(x)$
are the ground state and first excited stationary state of the system. 
($\phi_n(x)=\sqrt{2/a}$ $sin(n\pi x/a)$, $E_n=n^2 \pi^2 \hbar^2/(2ma^2)$ where $n=1,2,3...$)

\begin{center}
\epsfig{file=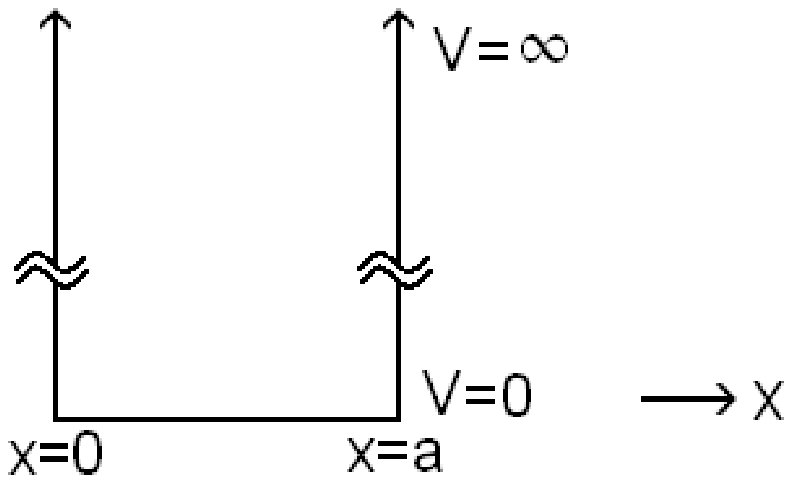,height=.8in}
\end{center}

Answer the following questions about this system:\\
(a) Write down the wave function $\Psi(x,t)$ at time $t$ in terms of $\phi_1(x)$ and $\phi_2(x)$.\\
(b) You measure the energy of an electron at time $t=0$. Write down the possible values of the energy and the probability 
of measuring each. \\
(c) Calculate the expectation value of the energy in the state $\Psi(x,t)$ above.\\
Table 1 shows percentage of correct responses for each part. 
The last column lists the percentage of students who displayed various kinds of common difficulties. 

\begin{table}[h]
\centering
\begin{tabular}[t]{|c|c|c|}
\hline
Q$\#$& $\%$Correct&Percentage of \\[0.5 ex]
& & Students with Common Difficulties \\[0.5 ex]
\hline
2 a &43 & 31$\%$ common phase factor, \\[0.5 ex]
 & & 9$\%$ no time-dependence\\[0.5 ex]
\hline
2 b&67 & 7$\%$ confusion between individual \\[0.5 ex]
& & measurement vs. expectation value\\[0.5 ex]
\hline
2 c&39 &17$\%$ wrote $\langle \Psi \vert E \vert \Psi \rangle$ \\[0.5 ex]
& &  or $\langle \Psi \vert H \vert \Psi \rangle$ but nothing else\\[0.5 ex]
\hline
\end{tabular}
\caption{Percentage of correct responses and the percentage of students with common difficulties on Q$\#$ (2).}
\label{junk2}
\end{table}
In question 2(a), we were expecting the response: \\
$\Psi(x,t)=\sqrt{2/7} \phi_1(x) e^{-iE_1t/\hbar}+\sqrt{5/7} \phi_2(x) e^{-iE_2t/\hbar}$. 
$43\%$ of the students provided correct responses. 
For tallying purposes, responses were considered correct if students wrote the phase factor 
for the first term as $e^{-iAE_1t}$ where $A$ is any real constant (e.g., $\hbar$ in the numerator, incorrect sign, or some other constant, 
e.g., mass $m$ in the phase were considered minor problems and ignored even though they can make the phase a quantity with dimension). 
Some students wrote incorrect intermediate steps; {\it e.g.},
$\Psi(x,t)=\Psi(x,0)e^{-iEt/\hbar}=\sqrt{2/7} \phi_1(x) e^{-iE_1t/\hbar}+\sqrt{5/7} \phi_2(x) e^{-iE_2t/\hbar}$. 
Such responses were considered correct for tallying purposes.
During the individual interviews, a student proceeded from an intermediate incorrect step to the correct time-dependence in the second step
similar to the expression above. Further probing showed that the student was having difficulty differentiating between the Hamiltonian 
operator and its eigenvalue and was probably thinking of
$\Psi(x,t)=e^{-i\hat H t/\hbar}\Psi(x,0)=\sqrt{2/7} \phi_1(x) e^{-iE_1t/\hbar}+\sqrt{5/7} \phi_2(x) e^{-iE_2t/\hbar}$ where
the Hamiltonian $\hat H$ acting on the stationary states gives the corresponding energies. 

As shown in Table 1, $31\%$ of students wrote common phase factors for both terms, {\it e.g.},
$\Psi(x,t)=\Psi(x,0)e^{-iEt/\hbar}$. Interviews suggest that these students were having difficulty differentiating
between the time-dependence of stationary and non-stationary states. Since the Hamiltonian operator governs the time-development
of the system, the time-dependence of a stationary state is via a simple phase factor. But non-stationary states in general
have a non-trivial time-dependence because each term in a linear superposition of stationary states evolves via a different phase factor.
Apart from using $e^{-iEt/\hbar}$ as the common phase factor, other common choices include $e^{-i \omega t}$, $e^{-i \hbar t}$,
$e^{-it}$, $e^{-ixt}$, $e^{-ikt}$ etc. 

Several students thought that the time dependence was a decaying exponential, 
e.g., of the type $\Psi(x,0)e^{-xt}$, $\Psi(x,0)e^{-Et}$,  $\Psi(x,0)e^{-ct}$, $\Psi(x,0)e^{-t}$ etc. During the interviews,
some of these students explained their choice by insisting that the wave function must decay with time.
Other incorrect responses were due to partial retrieval of related facts from memory such as the following:
\begin{itemize}
\item $\sqrt{2/7} \phi_1(x+\omega t)+\sqrt{5/7} \phi_2(x+\omega t) $
\item $\sqrt{2/7} \phi_1(x) e^{-i\phi_1 t}+\sqrt{5/7} \phi_2(x) e^{-i\phi_2 t}$
\item $\sqrt{2/7} \phi_1(x) e^{-ix t}+\sqrt{5/7} \phi_2(x) e^{-i 2 x t}$
\item $\sqrt{2/7} \phi_1(x) sin (\pi t)+\sqrt{5/7} \phi_2(x) cos(2 \pi t)$
\end{itemize}
Interviews suggest that these students often correctly remembered that the time-dependence of non-stationary states cannot be
represented by a common time-dependent phase factor but did not know how to correctly evaluate $\Psi(x,t)$. 
Interestingly, $9\%$ of the students wrote responses that did not have any time dependence.

In question 2(b), students were asked about the possible values of the energy of the electron
and the probability of measuring each in the initial state $\Psi(x,0)$. This turned out to be the easiest question on the survey. We expected students to
note that the only possible values of the energy in state $\Psi(x,0)$ are $E_1$ and $E_2$ and their respective probabilities are $2/7$ and $5/7$.
$67\%$ of students provided the correct response. $7\%$ got confused between individual measurements of the energy and its expectation value but
almost none of these students calculated the correct expectation value of the energy. Another common mistake was assuming that all allowed energies
for the infinite square well were possible and the ground state is the most probable because it is the lowest energy state.
Some students thought that the probabilities for measuring $E_1$ and $E_2$ are $4/(7a)$ and $10/(7a)$ respectively 
because they included the normalization factor for the stationary state wave functions $\sqrt{2/a}$ while squaring the coefficients.
Some believed that the probability amplitudes were the probabilities of measuring energy and did not square the coefficients
$\sqrt{2/7}$ and $\sqrt{5/7}$. Interviews suggest that students who were having difficulty with this question were very uncomfortable
with the poslutates of quantum mechanics and in making qualitative inferences from quantitative tools.

In question 2(c), students had to calculate the expectation value of the energy in the state $\Psi(x,t)$. The expectation value of the energy is
time-independent (it is the same in states $\Psi(x,t)$ and $\Psi(x,0)$)
because the Hamiltonian does not depend on time. If $\Psi(x,t)=C_1(t)  \phi_1(x)+C_2(t)  \phi_2(x)$, then the expectation
value of the energy in this state is $\langle E \rangle=P_1 E_1+P_2 E_2=\vert C_1(t)\vert^2 E_1+ \vert C_2(t)\vert^2 E_2=(2/7)E_1 +(5/7)E_2$
where $P_i=\vert C_i(t)\vert^2=\vert C_i(0)\vert^2$ is the probability of measuring the energy $E_i$. This question turned out to be difficult for students and
only $39\%$ provided the correct response. 

A surprising result is that a majority of students who had answered 2(b) correctly
did not see its relevance for 2(c), and did not exploit what they found for 2(b) in answering 2(c).
Consequently, many of the $39\%$ of students who answered question 2 (c) correctly 
worked out $\langle E \rangle$ from scratch by explicitly writing
 $\langle E \rangle=\int_{-\infty}^{+\infty}\Psi^*\hat H \Psi dx$, then writing
the wave function $\Psi(x,t)$ as a linear superposition of the ground state and first excited state, 
then using the fact that $\hat H$ acting on
a stationary state will give the corresponding energy and the same state back and using the orthonormalization of the wave function.
They explicitly showed that the time-dependent phase factors for the two terms that survive will go away due to complex conjugation. 
However, many students who tried to solve the problem this way by writing down the wave function explicitly inside the integral
got lost along the way. Some got lost early while others did not remember the orthonormalization condition or forgot to take
the complex conjugate of the wave function.
Some forgot that there was an integral involved in calculating the expectation value of a physical observable and wrote
$\langle E \rangle=\Psi^*\hat H \Psi$. Their final answers were in terms of the ground and first excited state wave functions.

Interviews suggest that there are two main reasons for why students did not exploit their response to part 2(b) for answering 2(c):
(I) Students did not recall the interpretation of expectation value as an ensemble average and thought that the only way
to calculate the expectation value is by the formal method above (during the interviews, these students incorrected claimed that
if the wave function is not given but the allowed energies and the probability of measuring each energy are given, it is not
possible to calculate $\langle E \rangle$), and (II) Students believed that the probability of measuring
different energies should depend on time (despite the fact that they evaluated $\Psi(x,t)$ in question 2 (a)).

In the interview, one student who answered 2(b) correctly did not know how to apply it to question 2(c). He wrote an explicit
expression involving the wave function for the ground and first excited states but thought that $\hat H\phi_n=E_n$ with no
$\phi_n$ on the right hand side of that equation. Therefore, he got a final expression for $\langle E \rangle$ that involved wave functions. When he was
told explicitly by the interviewer that the final answer should not be in terms of $\phi_1$ and $\phi_2$ and he should try to find his
mistake, the student tried his best but could not find his mistake. Then, the interviewer explicitly pointed to the particular step in which
he had made the mistake and asked him to find it. The student still had difficulty because he believed $\hat H\phi_n=E_n$ was correct. 
Finally, the interviewer told the student that
$\hat H\phi_n=E_n \phi_n$. At this point, the student was able to use orthonormality correctly to obtain the correct result
$\langle E \rangle=(2/7)E_1 +(5/7)E_2$. Then, the interviewer asked him to think whether it is possible to calculate
$\langle E \rangle$ based upon his response to 2 (b). The student's eyes brightened and he responded, ``Oh yes...I never thought of
it this way...I can just multiply the probability of measuring a particular energy with that
energy and add them up to get the expectation value because expectation value is the average value". Then, pointing to his detailed work for 
2(c) he added, ``You can see that the time dependence cancels out...".
$17\%$ of the students simply wrote $\langle \Psi \vert E \vert \Psi \rangle$ or $\langle \Psi \vert H \vert \Psi \rangle$ and did not 
know how to proceed.
A few students wrote the expectation value of energy as $[E_1+E_2]/2$ or $[(2/7)E_1 +(5/7)E_2]/2$.

\underline{Question (3):} 
Which of the following wave functions are allowed for an electron in a one dimensional infinite square well of width $a$
with boundaries at $x=0$ and $x=a$:
(I) $Asin^3(\pi x/a)$,\\
(II) $A[\sqrt{2/5} sin(\pi x/a)+ \sqrt{3/5} sin(2\pi x/a)]$ and\\
(III) $A e^{-{((x-a/2)/a)}^2}$? In each of the three cases, $A$ is
a suitable normalization constant. You must provide a clear reasoning for each case.

We hoped that  students would note that the last wave function $A e^{-{((x-a/2)/a)}^2}$ is not allowed because it does not satisfy the boundary 
conditions for the system (does not go to zero at $x=0$ and $x=a$). On the other hand, the first two wave functions 
$Asin^3(\pi x/a)$, $A[\sqrt{2/5} sin(\pi x/a)+ \sqrt{3/5} sin(2\pi x/a)]$ with suitable normalization constants $A$ are allowed.
They are both smooth functions that satisfy the boundary condition (each of them goes to zero at $x=0$ and $x=a$) so each 
can be written as a linear superposition of the stationary states. $79\%$ of students could identify that the second wave function is
 an allowed wave function because it is explicitly written in the form of a linear superposition of stationary states. 
Only $34\%$ gave the correct answer for all three wave functions. Within this subset,
a majority correctly explained their reasoning based upon whether the boundary conditions for the system are satisfied by these wave functions.
For tallying purposes, responses were considered correct even if the reasoning was not completely correct. 
For example, one student wrote (incorrectly): ``The first two wave functions are allowed because they satisfy the equation 
$\hat H \Psi=E \Psi$ and the boundary condition works". The first part of the reasoning provided by this student is incorrect while the
second part that relates to the boundary condition is correct.

$45\%$ believed that $Asin^3(\pi x/a)$ is 
not allowed but that $A[\sqrt{2/5} sin(\pi x/a)+ \sqrt{3/5} sin(2\pi x/a)]$ is allowed. 
Interviews suggest that a majority of students did not know that any smooth single-valued wave function that satisfies the boundary 
conditions can be written as a linear superposition of stationary states using the ``Fourier's trick". 
Interviews and written explanations suggest that many students incorrectly believed that the following two constraints 
must be independently satisfied for a wave function to be allowed:
(I) it must be a smooth single-valued function that satisfies the boundary conditions {\it and} 
(II) one must either be able to write it as a linear superposition of stationary states, or it must satisfy the Time-independent
Schroedinger equation (TISE).

As in the example below, some who correctly realized that $Asin^3(\pi x/a)$ satisfies the boundary condition, 
incorrectly claimed that it is still not an allowed state:
\begin{itemize}
\item $Asin^3(\pi x/a)$ satisfies b.c. but does not satisfy Schrodinger equation (i.e., it cannot represent a particle wave).
The second one is a solution to S.E. (it is a particle wave). The third does not satisfy b.c.
\end{itemize}

Many claimed that only pure sinusoidal wave functions are allowed, and $sin^2$ or $sin^3$ are not allowed.
Interviews and written explanations suggest that students believed that $Asin^3(\pi x/a)$ cannot be written as a 
linear superposition of stationary states and hence it is not an allowed wave function. The following are examples:
\begin{itemize}
\item $Asin^3(\pi x/a)$ is not allowed since it is not an eigen function nor a linear combination.
\item $Asin^3(\pi x/a)$ is not allowed because it is not a linear function but Schroedinger equation is linear.
\item $Asin^3(\pi x/a)$ is not allowed. Only simple sines or cosines are allowed.
\item $Asin^3(\pi x/a)$ works for 3 electrons but not one.
\end{itemize}

The most common incorrect response was claiming incorrectly that $Asin^3(\pi x/a)$ is not allowed because it
does not satisfy $\hat H\Psi=E \Psi$. 
Students asserted that $Asin^3(\pi x/a)$ does not satisfy TISE (which they believed was the equation that all allowed wave functions should 
satisfy) but $A[\sqrt{2/5} sin(\pi x/a)+ \sqrt{3/5} sin(2\pi x/a)]$ does.
Many explicitly wrote the Hamiltonian as $\frac{-\hbar^2}{2m} \frac{\partial^2}{\partial x^2}$ and showed that the second derivative of
 $Asin^3(\pi x/a)$ will not yield the same wave function back multiplied by a constant.
Incidentally, the same students did not attempt to take the second derivative of $A[\sqrt{2/5} sin(\pi x/a)+ \sqrt{3/5} sin(2\pi x/a)]$; otherwise 
they would have realized that even this wave function does NOT give back the same wave function multiplied by a constant. 
For this latter wave function, a majority claimed that it is allowed 
because it is a linear superposition of $sin(n\pi x/a)$. 
Incidentally, $Asin^3(\pi x/a)$ can also be written as a linear superposition of
only two stationary states. 
Thus, students used different reasoning to test the validity of the first two wave functions as in the following example:
\begin{itemize}
\item $\frac{-\hbar^2}{2m} \frac{\partial^2}{\partial x^2}Asin^3(\pi x/a)$ cannot be equal to $AEsin^3(\pi x/a)$ so it isn't acceptable. 
Second is acceptable because it is linear combination of sine.
\end{itemize}

Some students seemed completely confused. 
Some incorrectly noted that $A[\sqrt{2/5} sin(\pi x/a)+ \sqrt{3/5} sin(2\pi x/a)]$ is allowed inside the well 
and $A e^{-{((x-a/2)/a)}^2}$ is allowed outside the well.
Others incorrectly claimed that $Asin^3(\pi x/a)$ does not satisfy the boundary condition 
whereas $A[\sqrt{2/5} sin(\pi x/a)+ \sqrt{3/5} sin(2\pi x/a)]$ does. 
Some dismissed $Asin^3(\pi x/a)$ claiming it is an odd function that cannot be allowed for an infinite square well which is an even potential.
In the interview, a student who thought that only $A[\sqrt{2/5} sin(\pi x/a)+ \sqrt{3/5} sin(2\pi x/a)]$  is allowed said, ``these other
two are not linear superpositions". When the interviewer asked explicitly how he could tell that the other two wave functions cannot be written
as a linear superposition, he said, ``$Asin^3(\pi x/a)$ is clearly multiplicative not additive...you cannot make a cubic function out of 
linear superposition...this exponential cannot be a linear superposition either". 

$5\%$ of students claimed that $A e^{-{((x-a/2)/a)}^2}$ is
an allowed wave function for an infinite square well. These students did not examine the boundary condition. They sometimes
claimed that an exponential can be represented by sines and cosines and hence it is allowed or
focused only on the normalization of the wave function. 
Also, not all students who correctly wrote that
$A e^{-{((x-a/2)/a)}^2}$ is not allowed provided the correct reasoning. 
Many students claimed that the allowed wave functions
for an infinite square well can only be of the form $A sin(n\pi x/a)$ or that $A e^{-{((x-a/2)/a)}^2}$ is allowed 
only for a simple harmonic oscillator or a free particle.

\vspace*{-.1in}
\section{Summary}
\vspace*{-.1in}

The survey results and interviews indicate that students share common difficulties about basic quantum mechanics concepts, 
regardless of their background.
Instructional strategies that focus on improving student understanding of these concepts should take into account these difficulties.
We are currently developing and assessing Quantum Interactive Learning Tutorials (QuILT). 
QuILTs actively engage students in the learning process and help them build links between 
the abstract formalism and the conceptual aspects of quantum physics without compromising the technical content.

\vspace*{-.07in}
\begin{theacknowledgments}
We are grateful to the NSF for award PHY-0244708.
\end{theacknowledgments}

\bibliographystyle{aipproc}   

\end{document}